# Ultra-broadband Light Absorption by a Sawtooth Anisotropic Metamaterial Slab

Yanxia Cui[1, 2, 3], Kin Hung Fung[1,4], Jun Xu[1,4], Hyungjin Ma[1], Yi Jin[2],

Sailing He[2], and Nicholas X. Fang[1,4],*

[1] Department of Mechanical Science and Engineering and Beckman Institute of Advanced Science and Technology, University of Illinois at Urbana-Champaign, Urbana, Illinois 61801, USA

[2] Centre for Optical and Electromagnetic Research, State Key Laboratory of Modern Optical Instrumentation; Joint Research Centre of Photonics of the Royal Institute of Technology (Sweden) and Zhejiang University, Zhejiang University, Hangzhou 310058, China

[3] Department of Physics and Optoelectronics, Taiyuan University of Technology, Taiyuan 030024, China

[4] Department of Mechanical, Massachusetts Institute of Technology, Cambridge, Massachusetts 02139, USA

*Corresponding author:nicfang@mit.edu

## Abstract

We present an ultra broadband thin-film infrared absorber made of saw-toothed anisotropic metamaterial. Absorbtivity of higher than 95% at normal incidence is supported in a wide range of frequencies, where the full absorption width at half maximum is about 86%. Such property is retained well at a very wide range of incident angles too. Light of shorter wavelengths are harvested at upper parts of the sawteeth of smaller widths, while light of longer wavelengths are trapped at lower parts of larger tooth widths. This phenomenon is explained by the slowlight modes in anisotropic metamaterial waveguide. Our study can be applied in the field of designing photovoltaic devices and thermal emitters.

Metamaterials (MMs) are artificial materials engineered to exhibit extraordinary electromagnetic properties that are not available in nature.[1,2] Potential applications of MMs are diverse and include superlenses,[3] invisible cloaks,[4] highly sensitive sensors,[5] ultrafast modulators,[6] antenna systems.[7] In most applications, the absorption loss in MMs often degrades the performance; however, for artificial light absorbers, the absorption loss becomes useful and can be significantly enhanced by proper designs of MMs.[8,9]

In 2008, Landy *et. al* proposed a single-wavelength perfect absorber consisting of metallic split ring resonators and cutting wires.[10] Later, some improvement works were followed to make the absorbers insensitive to incident angle and polarization.[11,12] Unfortunately, those past efforts suffer common disadvantage of narrow bandwidth, which will reflect a fairly large amount of total incident energy and could not be employed to adequately improve the solar energy harvesting efficiency.[8,9] One may suggest adding various different resonances in order to broaden the absorption band, [13-20] but the strong coupling among resonators often put more limitations so that the designed absorbers often perform much poorer in comparison with a black body which absorbs all incident electromagnetic radiation. We note that Yang *et. al* have designed an ultra-broadband absorber based on an array of metallic nanogrooves of different depths.[21] However, it is almost impractical to obtain metallic grooves with parameters like 10 nm width and 5 $\mu$m depth using current fabrication technologies. Therefore, MM structures of simple schematics which can absorb light efficiently in a broadband are very demanding.

In this letter, based on slowlight waveguide modes of weakly coupled resonances in a MM slab, we design an ultra-broadband thin film infrared absorber for TM-polarized light. The slowlight waveguide can be obtained by etching a MM slab into sawtooth shape with the tooth widths increasing gradually from top to bottom, Figure 1. Our broadband absorber can be regarded as a group of ultra-short vertical waveguides which support slowlight modes at different frequencies so that the incident light at different wavelengths can be captured at positions of different tooth widths. In detail, we employ an anisotropic MM (AMM) consisting of alternating layers of flat metal and dielectric plates. Metal plates are made of gold with thickness $t_m$ = 15 nm; dielectric plates are made of germanium with thickness $t_d$ = 35 nm. The permittivity of gold ($\varepsilon_m$) are from Ref. 22, and the relative permittivity of germanium is $\varepsilon_d$ = 16. The total number of metal/dielectric pairs (*N*) is 20. The slab is carved into periodically spaced sawteeth with period *P* = 800 nm, top width $W_s$ = 150 nm and bottom width $W_l$ = 600 nm. A gold film with a thickness (*t* = 100 nm) large enough is added under the sawtooth AMM slab to block all transmission.

Using the Rigorous Coupled-wave Analysis (RCWA) method, [23,24] we simulate a plane wave of TM polarization (magnetic field $H_y$ perpendicular to the *x-z* plane) impinging on the structure. The absorbtivity (defined by $\eta = 1-R-T$, where *T* = 0) are calculated based on the Poynting theorem. The obtained absorption spectrum at normal incidence (Figure 2a, Thick) indicates that the absorption performance is excellent, with absorbtivity higher than 95% covering the range from 3 to 5.5 μm and a full absorption width at half maximum (FWHM) of 86%. It is important to note that most of the MM absorbers have very narrow absorbing bands; for instance, the FWHM of the MM absorber based on metallic crosses in Ref. 25 is only 16.7%. In comparison, the present absorber has an ultra-broad absorbing band, which is about 5 times that of a traditional single band absorber. To evaluate its overall capability of light conversion for applications of light harvesting and optical detectors, it is also meaningful to calculate the total absorbed energy efficiency which is the integration of absorption over the total incident energy at the regarded energy band, i.e., $\Delta = \int_{\omega_1}^{\omega_2} \eta(\omega)d\omega/(\omega_2 - \omega_1)$. Here the energy band corresponding to the regarded wavelength range from 2.5 to 7 *μm* is considered. Our AMM absorber has *Δ* = 86.4%, which reflects that most of the incident energy within the considered range can be absorbed. Our absorber also performs very well in a wide angular range. The absorption map with incident angles ($\phi$) versus wavelengths is plotted in Figure 2b with a curve indicating the contour of $\eta$ = 0.9. It is seen that when the incident angle $\phi$ is smaller than 60 degrees, the structure keeps the performance of very high and ultra-broadband absorption, resembling that at normal incidence.

The metal/dielectric sawteeth can be described as metamaterials with effective constitutive parameters.[26] In this case, there are two effective permittivities, $\varepsilon_\parallel$ and $\varepsilon_\perp$, for the electric-fields along directions parallel and perpendicular to the metal/dielectric interfaces, respectively. It is known that $\varepsilon_\| = f\varepsilon_m(\omega) + (1-f)\varepsilon_d$, $\frac{1}{\varepsilon_\perp} = \frac{f}{\varepsilon_m(\omega)} + \frac{1-f}{\varepsilon_d}$, where $f = \frac{t_m}{t_m + t_d}$ is the filling ratio of the metal.[1] Therefore, each metal/dielectric sawtooth can be considered as a sawtooth made of homogeneous material with anisotropic permittivities, as shown in the inset of Figure 2a. We find that the effective sawtooth structure has an absorption spectrum (Figure 2a, Thin) very similar to the actual sawtooth structure (Figure 2a, Thick). Such agreement illustrates that the effective medium description used here is valid.

To understand how light is absorbed in our sawtooth AMM absorber, we study the normalized field distributions $|H_y|$ in the actual inhomogeneous sawtooth structure at three different wavelengths; see the color maps in Figure 3 with $\lambda_0$ = 3.5, 4.5, and 5.5 $\mu$m, respectively. It is identified that light of different wavelengths accumulate at different parts of the sawtooth absorbers (indicated by the dotted arrow in each subfigure). For instance, light of a shorter wavelength at $\lambda_0 = 3.5\mu$m (Figure 3a) is harvested at the upper part of the sawteeth of a smaller width ($W$ = 277.1 nm) and the field amplitude at the bottom part has very small values. However, for the incident light of a longer wavelength at $\lambda_0 = 5.5\mu$m (Figure 3c), the energy is trapped at the lower part of the AMM slab where the sawteeth have a larger width ($W$ = 479.6 nm), whereas around the top parts there is almost no strong field concentration. Different from Figures 3a and 3c, light at a wavelength close to the middle of the absorbing band of $\lambda_0 = 4.5\mu$m (Figure 3b) is focused at the middle waist region with $W$ = 389.6 nm. It is noted that for the effective homogeneous sawtooth structure shown in the inset of Figure 2a, the calculated field distributions are almost the same except some details around metal/dielectric interfaces. It is also noted that in Figure 3 the field amplitude within the metal thin film plates around the concentration centre is very high (though a little lower than that in the dielectric plates). This is because the thickness ($t_m$) of the metal plates is close to the skin depth of metal within the studied frequency range which allows light directly penetrating through the metal plates. In addition, we note that it is unfeasible to utilize a model of very small number to discretize the effective medium slab.

Plots of Poynting vector (**S**) (Figure 3, arrow maps) can indicate how the light propagates in the AMM absorber before it is totally absorbed. One sees that most of the incident energy first propagates downwardly along the *z* direction in the air gap region without penetrating into the AMM sawteeth and then whirls into the AMM region, forming vortexes close to the interface between AMM and air regions. The vortexes locate right at the places where the magnetic field concentrates. These are typical features in slowlight waveguides [27-29] with slowlight defined as the propagation of light at a very low group velocity in comparison with light speed in vacuum. For example, in Ref. 27 the authors demonstrated that the designed air/dielectric/metal waveguide could support such kind of vortex mode propagating very slowly along the slab and even being trapped at a critical core thickness. Later, an AMM/air/AMM slowlight waveguide has also been presented to slow down the electromagnetic propagating waves to a complete standstill at certain regions of the air core waveguide.[28]

Similar to the references above, we can also employ the concept of slowlight to understand the physical principle of absorption in our structure. The configuration of the three-layered air/AMM/air waveguide (Figure 4a) has the AMM core of width *W*. Based on Maxwell equations and boundary conditions, its dispersion relationship between the incident photon frequency ($\omega_c = \omega/c$) and the propagating constant (*β*) is derived as

$$\exp[iq_2 W] + \frac{\frac{\kappa_1}{n_0^2} - i\frac{q_2}{\varepsilon_\perp}}{\frac{\kappa_1}{n_0^2} + i\frac{q_2}{\varepsilon_\perp}} = 0, \tag{1}$$

where $\kappa_1 = \sqrt{\beta^2 - n_0^2 \omega_c^2}$, $q_2 = \sqrt{\varepsilon_\perp \omega_c^2 - \frac{\varepsilon_\perp}{\varepsilon_\parallel}\beta^2}$, and $n_0$ is the refractive index of air. For air/AMM/air waveguide of different fixed core widths, their dispersion diagrams ($\omega_c$-*β*) can be calculated. Dispersion curves of fundamental mode at *W* = 3, 4, 5, and 6 *μ*m are shown in Figure 4b. One can observe that for each dispersion curve, at lower frequency band the propagating constant increases gradually with the incident photon frequency; however, when the incident light approaches the cut-off frequency, the dispersion curve becomes flat. A close-up view demonstrates that these dispersion curves are not strictly flat around the cut-off frequencies but declines slightly after passing them. Figure 4c shows that the fundamental propagating mode has an extreme point with $\frac{|v_g|}{c}$ approaches 0 (i.e., group velocity $v_g \equiv \frac{d\omega_c}{d\beta} = 0$). This reflects that slow light modes are supported in the air/AMM/air waveguide, similar to those mentioned in Refs. 27-29. Therefore, for a waveguide of a certain core width, the slowlight modes can be generated around a certain photon frequency (or wavelength). Photon wavelength points at $v_g = 0$ for different waveguide core widths (*W*) (Figure 4d, solid squares) indicate that wavelength of light exciting the slow light modes increases linearly when *W* increases. This can be explained as follows. Considering the practical material parameters, as the photon frequency is close to the cut-off frequency band, Eq. (1) can be simplified by using $-\exp[iq_2W] \approx 1$. For the fundamental mode, further simplification gives that $\frac{\omega}{c} = \frac{\pi}{W\sqrt{\varepsilon_\perp}}$, based on which the relationship between the waveguide core width and the incident light wavelength exciting the slowlight mode can be write approximately in a very simple form

$$\lambda_p \approx 2W\sqrt{\varepsilon_\perp} \qquad (2).$$

Taking *W* = 400 nm as an example with the effective permittivity $\varepsilon_\perp$ = 23.1+0.05*i*, we obtain that $\lambda_p$ equals 3.8 *μ*m, which refers to the corresponding numerical point in Figure 4d. From Eq. (2), it can also be known that the effective refractive index of the air/AMM/air is $\sqrt{\varepsilon_\perp}$ which can be tuned not only by adjusting the material properties ($\varepsilon_d$) as a traditional waveguide but also by varying the filling ratio of the metal/dielectric composite plate (*f*). Therefore, there is more freedom to tune the absorption spectrum.

Due to the evidences mentioned above, we can regard our absorber as a group of air/AMM/air waveguide absorbers of infinite number, which can respond at different frequencies and then display as a collective effect. Figure 4d suggests that the air/AMM/air waveguide of *W* from 150 nm to 600 nm (the range of sawteeth width of our AMM absorber) can support the slow light mode from wavelength 2 to 6 $\mu$m, which matches nicely with the high absorption band of our absorber as shown in Figure 2a. Last, we compare the relation between the slowlight wavelength and AMM widths obtained from the dispersion relation (Figure 4b) with that between the incident wavelength and the AMM widths at the locations of magnetic field maxima in the electromagnetic field simulation (Figure 3). Figure 4d shows a good agreement between the two relation curves. The deviation is rational because the width of the AMM core is fixed in the dispersion calculation while our absorber in Figure 1 has a gradually tuned core width. If the field maximum is closer to the bottom, the deviation becomes larger due to the additional impact from the bottom metallic reflector. Based on all these properties, we conclude that the present ultra-broadband absorber is based on the slowlight modes supported by the air/AMM/air waveguides.

Although slowlight waveguides have been proposed before, the present work is pioneering in its application to broadband thin film absorbers. Because the slowlight modes have weakly coupled resonances, we can construct broadband absorbers by gradually tuning the width of the waveguide core; since the modes have local resonances in subwavelength scale, the absorption is almost angular independent. Additionally, the absorption spectrum can be easily tuned by changing the filling ratio of metal. Moreover, we can easily extend it to 3D to achieve a corresponding AMM surface corrugated with periodical pyramids which works polarization-independently. The weakness of the present absorber is the small absorption cross section compared with traditional metamaterial absorbers[15] because it is based on weakly coupled resonances. Thus, if we increase the pitch of the AMM sawteeth, the efficiency of the absorber will decrease.

In conclusion, we have designed an ultra-broadband and wide-angle thin film absorber based on an anisotropic metamaterial which has been corrugated into sawtooth shape. The frequency absorption band with higher than 95% absorbing efficiency has a FWHM of 86% at normal incidence, which is around 5 times of a traditional single band absorber. This absorber works very well in a wide angular range, from 0 to 60 degrees. The total thickness of the film slab is around a quarter of the central wavelength of the absorption band. The absorber captures light of different incident wavelengths at positions of different tooth widths according to the slowlight principle of a three-layered air/AMM/air anisotropic waveguide. It is expected that our study can be applied in the field of designing photovoltaic devices and thermal emitters of very good performance. We emphasize that our proposal is scalable and can be applied at microwave and terahertz frequencies as well.

**Acknowledgements** This work is partially supported by the National Science Foundation (CMMI 0846771), the National Natural Science Foundation of China (60990320 and 60901039) and AOARD (114045). Cui acknowledges the research foundation of Taiyuan University of Technology.

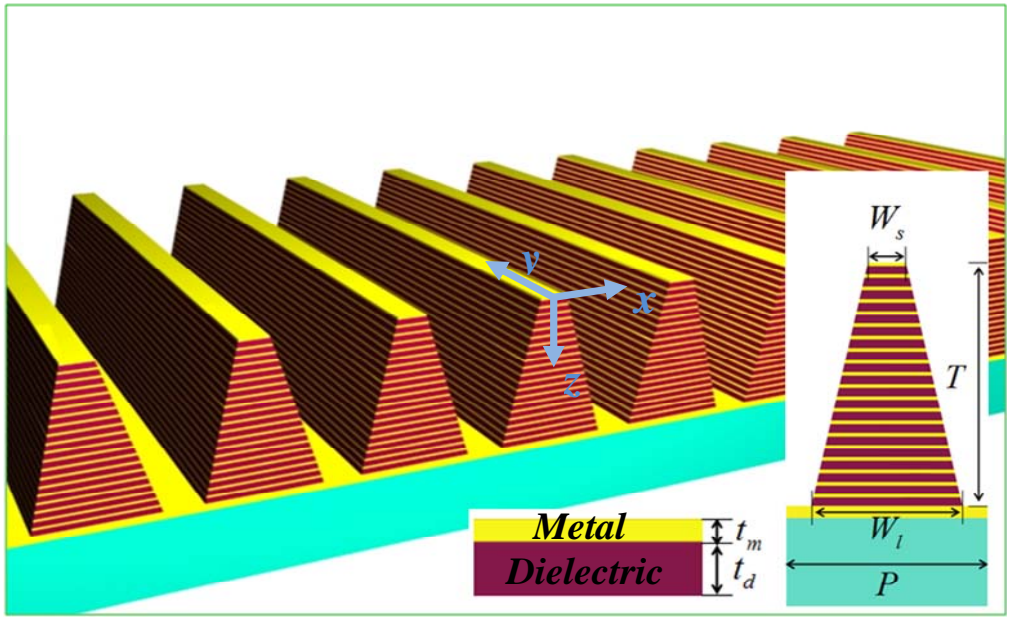

**Figure 1** Diagram of the sawtooth anisotropic metamaterial thin film absorber. $P$ = 800 nm, $T$ = 1000 nm, $W_s$ = 150 nm, $W_l$ = 600 nm, $t_d$ = 35 nm, and $t_m$ = 15 nm.

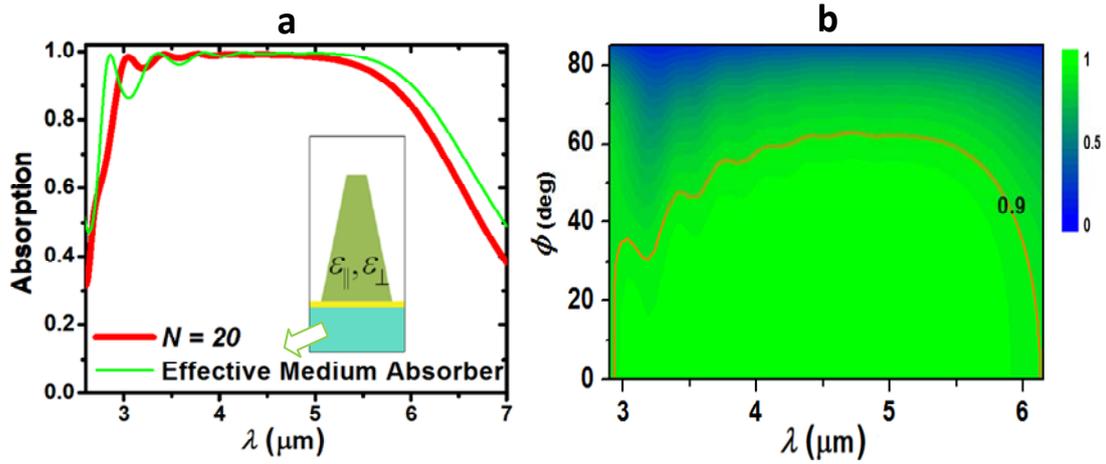

**Figure 2** (a) Absorption spectra for the sawtooth AMM absorber (Thick) with number of periods $N = 20$ and the effective homogeneous sawtooth structure that is shown in the inset of Figure 2a (Thin). (b) Angular absorption spectrum of the sawtooth AMM film in Figure 1; the line represents the efficiency contour with $\eta = 0.9$.

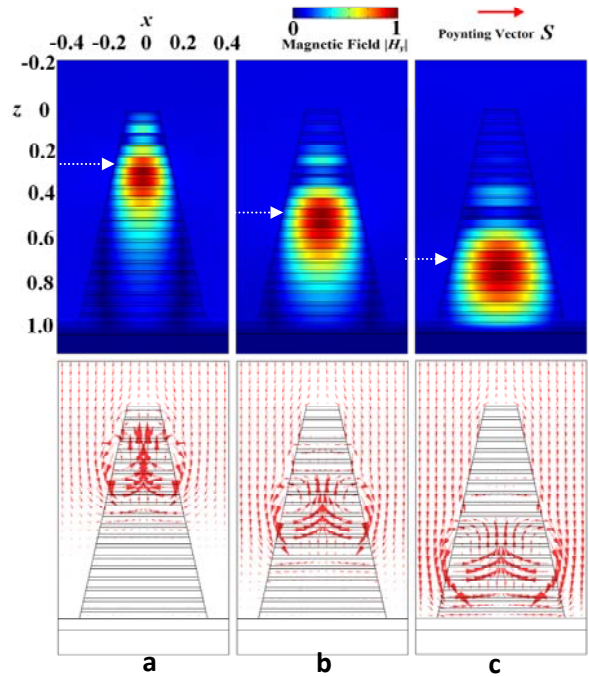

**Figure 3** Distributions of magnetic field (color maps) and energy flow (arrow maps) for the sawtooth AMM absorber at different incident wavelengths: (a) $\lambda_0 = 3.5\ \mu m$, (b) $\lambda_0 = 4.5\ \mu m$, and (c) $\lambda_0 = 5.5\ \mu m$. The vertical positions of field maxima are indicated with dotted arrows in the magnetic field plots.

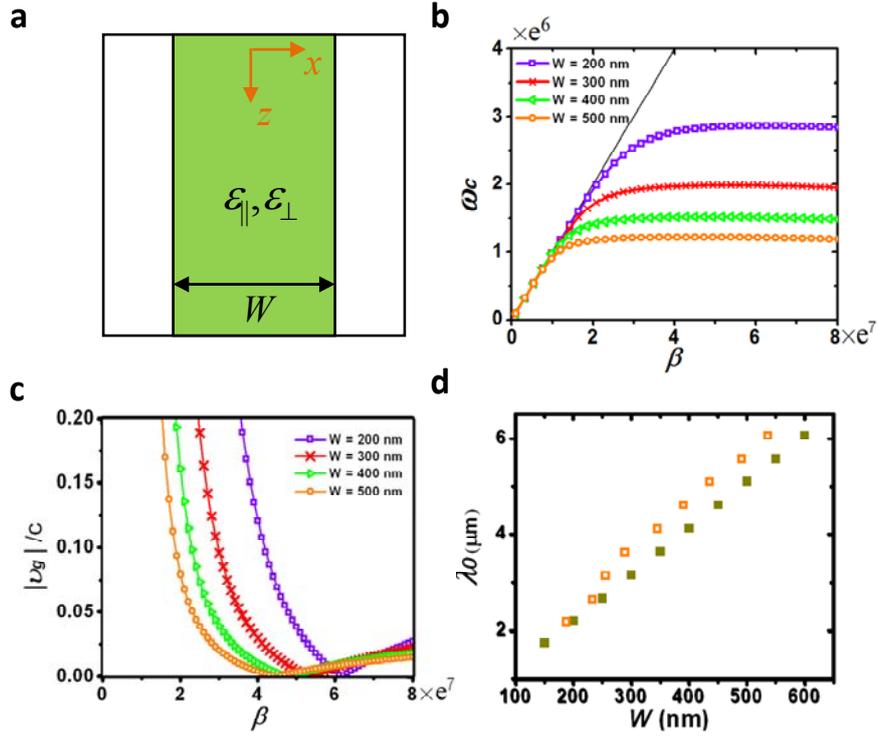

**Figure 4** (a) Diagram of the air/AMM/air waveguide with a fixed core width of $W$; (b) dispersion curves of the air/AMM/air waveguides at different $W$; (c) relationship between the group velocity and wavevectors at different $W$; (d) Solid square points: the wavelength at $v_g = 0$ at different $W$, Hollow square points: the width of the AMM teeth at the energy centre as plotted in Figure 3 for different wavelengths.